\documentclass[12pt]{article}
\usepackage[dvips]{graphicx}

\usepackage{graphics,epsfig}

\renewcommand{\bar}[1]{\overline{#1}}
\begin{document}

\begin{center}
{\Large
\bf Mass of $D_{sJ}^*$(2317) and Coupled Channel Effect}\\

%\vspace{13pt}

\vspace{1.5cm}

Dae Sung Hwang${}^{(a)}$ and Do-Won Kim${}^{(b)}$\\
{\it{a: Department of Physics, Sejong University, Seoul 143--747,
Korea}}\\
{\it{b: Department of Physics, Kangnung National University,
Kangnung 210-702, Korea}}\\

\vspace{2.5cm}

{\bf Abstract}\\
\end{center}
\noindent
The resonance $D_{sJ}^*$(2317) which is considered to be the
$1{}^3P_0$ state composed of charm and strange quarks has been discovered recently.
The measured mass, which is about 160 MeV lower than the
mass of the $1{}^3P_0$ state obtained from the potential model calculation by
Godfrey and Isgur, was considered surprisingly low and attracted a lot of
theoretical investigations.
We calculate the mass shift of the $1{}^3P_0$ state by using
the coupled channel effect. Our result shows that the coupled channel effect
naturally explains the observed mass of $D_{sJ}^*$(2317).
\\

\vfill


\noindent
$^a$e-mail: dshwang@sejong.ac.kr\\
$^b$e-mail: dwkim@kangnung.ac.kr\\
\thispagestyle{empty}
\pagebreak

\setlength{\baselineskip}{13pt}

\section{Introduction}

The BaBar Collaboration \cite{babar03} recently discovered a narrow resonance
in $D_s^+\pi^0$, which is known as the $D_{sJ}^*(2317)$,
and it was confirmed by the CLEO \cite{cleo03}
and the Belle Collaborations \cite{belle03}.
Its decay patterns suggest a quark-model $0^+$ classification.
The measured mass, 2317.4$\pm$0.9 MeV \cite{pdg04} which is 40.9$\pm$1.0
MeV below the threshold of $D^0K^+$, was considered surprisingly low compared to
the predictions of the potential model calculations.
For example, the prediction of the $1{}^3P_0$ mass by
Isgur and Godfrey \cite{IG} was 2.48 GeV,
and that by Eichten and Di Pierro \cite{EP} was 2.487 GeV,
which are about 160 and 170 MeV higher than the measured mass of $D_{sJ}^*(2317)$.

There have been many theoretical investigations which aimed to explain the measured
low mass of $D_{sJ}^*(2317)$.
Cahn and Jackson \cite{CJ} used effective spin-orbit and tensor forces resulting from
a model based on the nonrelativistic reduction.
Bardeen et al. \cite{BEH} worked with effective Lagrangians based on the chiral symmetry
in heavy-light meson systems.
Barnes et al. \cite{BCL} considered a mixing between two molecular states
$|D^0K^+>$ and $|D^+K^0>$.
Browder et al. \cite{BPP} proposed a mixing between the $q{\bar{q}}$ and 4-quark states
and assigned a linear combination with less mass as $D_{sJ}^*(2317)$.

In this paper we calculate the mass shift of the $1{}^3P_0$ state composed of
$c$ and $\bar{s}$ quarks by using the coupled channel effect which was pioneered
by Eichten et al. \cite{eichten12}.
We consider the coupling of the $1{}^3P_0$ bound state, 
whose bare mass is 2.48 GeV given in \cite{IG}, with the two-meson
continuum sectors of $DK$, $D^*K$, $DK^*$ and $D^*K^*$.
In reality, the $D^*K$ and $DK^*$ do not contribute since their
spin-dependent statistical coupling coefficients are zero.
We find that the contribution to the transition amplitudes from the Coulombic
part of the interaction potential between quarks
is about 60 \% of that from the linear part.
This aspect was first realized by Zambetakis \cite{zam1,byers94},
and it is contrary to the usual consideration that the Coulombic part is negligible
in the contribution to the transition amplitudes \cite{eichten12,ELQ}.
As a consequence, the magnitude of the reduced interaction hamiltonian
resulting when both linear and Coulombic parts are included is about 2.6
times that resulting when only the linear part is considered.
This fact is crucial in the analysis of this paper:
When the Coulombic part is not included, there is no physical mass eigenstate
below the threshold of $D^0K^+$ for physically acceptable parameterizations.
On the other hand, when the Coulombic part is included, there exists a
physical mass eigenstate with its mass eigenvalue close to the measured $D_{sJ}^*(2317)$
mass in a natural manner.
This paper is organized as follows. In section 2 we briefly summarize the
coupled channel effect.
Even though it is explained in detail in Refs. \cite{eichten12,zam1,byers94},
we think that a concise summary is useful for this paper.
In section 3 we explain the procedures and results of our calculation.
The last section is conclusion.

\section{Coupled Channel Effect}

\subsection{Basics}

The coupled channel effect takes into account the fact that
two-meson continuum states of $DK$ couple to two-quark $c{\bar{s}}$
bound states.
There are two important effects of this coupling.
The first is to give rise to mass shifts and configuration mixing of
$c{\bar{s}}$ bound states.
The second is to provide broad peaks of $c{\bar{s}}$ resonance
states above threshold \cite{eichten12,zam1,byers94}.
In this section we briefly summarize the coupled channel effect
which has been explained in detail in Refs. \cite{eichten12,zam1,byers94}.

When we take the coupled channel effect into consideration,
Hilbert space is composed of two sectors:
discrete $c{\bar{s}}$ bound states and continuum states of two mesons.
Then the total hamiltonian is written as
\begin{equation}
H=H_0+H_{PC},\  \ \ {\rm where}\ \ \
H_0=\left(\begin{array}{cc}
H_Q&0\\
0&H_C\\
\end{array}
\right) ,
\ \ \
H_{PC}=\xi\left(\begin{array}{cc}
0&H_{QC}\\
H_{CQ}&0\\
\end{array}
\right) .
\label{a1}
\end{equation}
$H_Q$ and $H_C$ are diagonal respectively in the quark-antiquark bound state
space $|i>$ and in the two-meson continuum space $|{\bf P},\lambda >$
(where ${\bf P}$ is the relative momentum of two mesons),
which satify
\begin{equation}
H_0|i>=M^{\rm bare}_i|i>,\ \ \
H_C|{\bf P},\lambda > = E_C(P)|{\bf P},\lambda >\ ,
\label{a2}
\end{equation}
%where
\begin{equation}
<i|j>=\delta{ij},\ \ \
<{\bf P},\lambda | {\bf P}',\lambda' >=(2\pi)^3\delta^3({\bf P}-{\bf P}')
\delta_{\lambda \lambda'},\ \ \
<i| {\bf P},\lambda >=0 .
\label{a3}
\end{equation}
$H_{PC}$ in (\ref{a1}), where $PC$ means pair creation, connects two spaces
$|i>$ and $|{\bf P},\lambda >$.

Physical mass eigenstates are eigenstates of the total hamiltonian $H$ in (\ref{a1}).
We denote the physical mass eigenstate by $|N>$ with a collective quantum number $N$.
Then with a physical mass $M_N$, $|N>$ satisfies
\begin{equation}
H|N>=M_N|N>.
\label{a4}
\end{equation}
When the value of $\xi$ in (\ref{a1}) varies to zero, $|N>$ becomes one of the bare
states $|i>$.
Physical mass eigenstate $|N>$ can be expanded in terms of bare states as
\begin{equation}
|N>=\sum_i a_i|i>+\sum_{\lambda}\int {d^3P\over (2\pi)^3}b_{\lambda}(P)
|{\bf P},\lambda >.
\label{a5}
\end{equation}
{}From (\ref{a1})-(\ref{a5}) $b_{\lambda}(P)$ and $a_i$ in (\ref{a5}) are
related by
\begin{equation}
b_{\lambda}(P)=\sum_i{<{\bf P},\lambda | H_{PC} |i>\over M_N-E_C(P)}a_i ,
\label{a6}
\end{equation}
then from (\ref{a1}) and (\ref{a4}) we get
\begin{equation}
H_{\rm eff}a^N=M_Na^N\ \ \ {\rm with}\ \ \
H_{\rm eff}=M^{\rm bare}+\Omega(M_N),
\label{a7}
\end{equation}
where $M^{\rm bare}$ is a diagonal matrix with $M^{\rm bare}_i$ in
(\ref{a2}) as diagonal elements, $a^N$ is a column vector with probability amplitudes
$a^N_i$ as its elements, and the matrix elements of $\Omega(W)$ is given by
\begin{equation}
\Omega_{ij}(W)=\sum_{\lambda}\int {d^3P\over (2\pi)^3}
{<i|H_{PC}| {\bf P},\lambda ><{\bf P},\lambda | H_{PC} |j> \over W-E_C(P)+i\epsilon} .
\label{a8}
\end{equation}
For $W$ above threshold, $\Omega (W)$ has the structure of
$\Omega (W)= \Delta (W) - i\Gamma (W)/2$ with $\Delta (W)$
and $\Gamma (W)$ real symmetric
matrices.
Once the matrix $\Omega (W)$ is known, the mass eigenvalues of the
coupled system are found from
\begin{equation}
{\rm Det}(M_NI-[M^{\rm bare}+\Omega(M_N)])=0.
\label{a9}
\end{equation}

%For the existence of non-trivial solutions of (\ref{a7}) 

We can calculate the normalization constant of the amplitudes in (\ref{a5})
from
\begin{equation}
{\cal N}^2\Bigl(\sum_i|a_i|^2 + \sum_{\lambda}\int {d^3P\over (2\pi)^3}
|b_{\lambda}(P)|^2\Bigr)
=1.
\label{a10}
\end{equation}
Using (\ref{a6}) and (\ref{a8}), we get
\begin{equation}
{\cal N}^2=\Bigl(\sum_i|a_i|^2 + \sum_{i,j} a_i^* \omega_{ij}(M_N)a_j
{\Bigr)}^{-1} ,
\label{a11}
\end{equation}
where
\begin{equation}
\omega_{ij}(M_N)=-{d\over dW}\Omega_{ij}(W)|_{W=M_n} .
\label{a12}
\end{equation}
The probability that the physical mass eigenstate $|N>$ be in the quark-antiquark
bound state sector is given by $Z={\cal N}^2\sum_i|a_i|^2$, while that in
the two-meson state sector is $1-Z$.

\subsection{Mechanism of Light Quark Pair Creation}

The Cornell group studied the effect of OZI allowed decay channels \cite{eichten12}.
They proposed that the following interaction hamiltonian is responsible for the decay
as well as the binding of quark--antiquark bound states.
\begin{equation}
H_I = {1\over 2} \sum_{a=1}^8 \int d^3x d^3y : \rho_a({\bf x})
V({\bf x}-{\bf y})\rho_a({\bf y}) : ,
\label{a13}
\end{equation}
where
\begin{equation}
V(r)=-{\kappa\over r}+{r\over a^2}.
\label{a14}
\end{equation}
In (\ref{a13}) $\rho_a({\bf x})=\psi^{\dagger}({\bf x}){1\over 2}\lambda_a\psi({\bf x})$
are the color densities of quark fields.
This model corresponds to the vector coupling since (\ref{a13}) is the leading term of
the vector coupling hamiltonian in the nonrelativistic expansion.

One can calculate the transition amplitude between two-quark bound and two-meson
continuum states using the interaction hamiltonian (\ref{a13}).
Then (\ref{a8}) becomes
\begin{eqnarray}
\Omega_{nL,mL'}(W)&=&\sum_I\int P^2dP
{H^I_{nL,mL'}(P)\over W-E_1(P)-E_2(P)+i\epsilon}
\nonumber\\
&\equiv&
\sum_I[\Delta^I_{nL,mL'}(W)-{i\over 2}\Gamma^I_{nL,mL'}(W)] ,
\label{a15}
\end{eqnarray}
where
\begin{equation}
H^I_{nL,mL'}(P)=f^2\sum_l C(Js,LL',J_1J_2,l)
I_{nL}^l(P) I_{mL'}^l(P) 
\label{a16}
\end{equation}
with
\begin{equation}
f^2={2\over 3\pi^2 a^4 m_q^2} \, {1\over \beta^3}.
\label{a17}
\end{equation}
In (\ref{a16}), $I_{nL}^l(P)$ is the momentum dependent factor of the transition amplitude.
The superscript $I$ in (\ref{a15}) denotes the participating coupled channels
which are composed of two mesons.
The angular momenta and energies of these two mesons are denoted by $J_1$, $J_2$, and
$E_1(P)$, $E_2(P)$, respectively.
$H^I_{nL,mL'}(P)$ depends on the coupled channel, and it also depends on $J$ and $s$
(the total angular momentum and spin of the quark-antiquark bound state).

For the vector coupling, $I_{nL}^l(P)$ in (\ref{a16}) is given by
\begin{equation}
I_{nL}^l(P)=\int_0^{\infty}dt \Theta (t) R_{nL} ({t\over {\sqrt{\beta}}})
j_l({\rho_Q P t \over {\sqrt{\beta}}}) 
\label{a18}
\end{equation}
with
\begin{equation}
\Theta (t)=[te^{-t^2}+(t^2-1)e^{-t^2/2}{\sqrt{\pi\over 2}} erf({t\over {\sqrt{2}}})]
+4\beta a^2 \kappa [-te^{-t^2}+e^{-t^2/2}{\sqrt{\pi\over 2}} erf({t\over {\sqrt{2}}})] ,
\label{a19}
\end{equation}
where
$R_{nL}(r)$ is the radial wave function of the quark-antiquark bound state,
$j_l(t)$ is the spherical Bessel function,
and $\rho_Q = m_Q/(m_q+m_Q)$.
The first and second terms in (\ref{a19}) come from the linear and Coulombic parts
of (\ref{a14}), respectively.

\section{Mass Shift of $D_{sJ}^*$ (2317) Meson from Coupled Channel Effect}

Isgur and Godfrey obtained the mass of the $1{}^3P_0$ state composed of $c$
and ${\bar{s}}$ quarks as 2.48 GeV from the interaction between two quarks.
Their result corresponds to the bare mass $M^{\rm bare}$ in (\ref{a2}).
In this section we calculate the mass shift of the $1{}^3P_0$ state using
the coupled channel effect explained in the previous section.
In particular, (\ref{a9}) gives the physical mass given when the
coupled channel effect is incorporated.

\begingroup
%\squeezetable
\begin{table}[t]
%\vspace*{0.3cm}
\label{tab:mesonmass}
\begin{center}
\begin{tabular}{cccccccc}
\hline\hline
$D^0$&$D^+$&$D^{*0}$&$D^{*+}$&
$K^+$&$K^0$&$K^{*+}$&$K^{*0}$\\
\hline
1864.6&1869.4&2006.7&2010.0&
493.677&497.648&891.66&896.10\\
\hline\hline
\end{tabular}
\end{center}
\vspace*{-0.5cm}
\caption{Meson Masses (MeV) used in the calculation \cite{pdg04}.}
\end{table}
\endgroup

\begin{figure}
\centering
\epsfig{figure=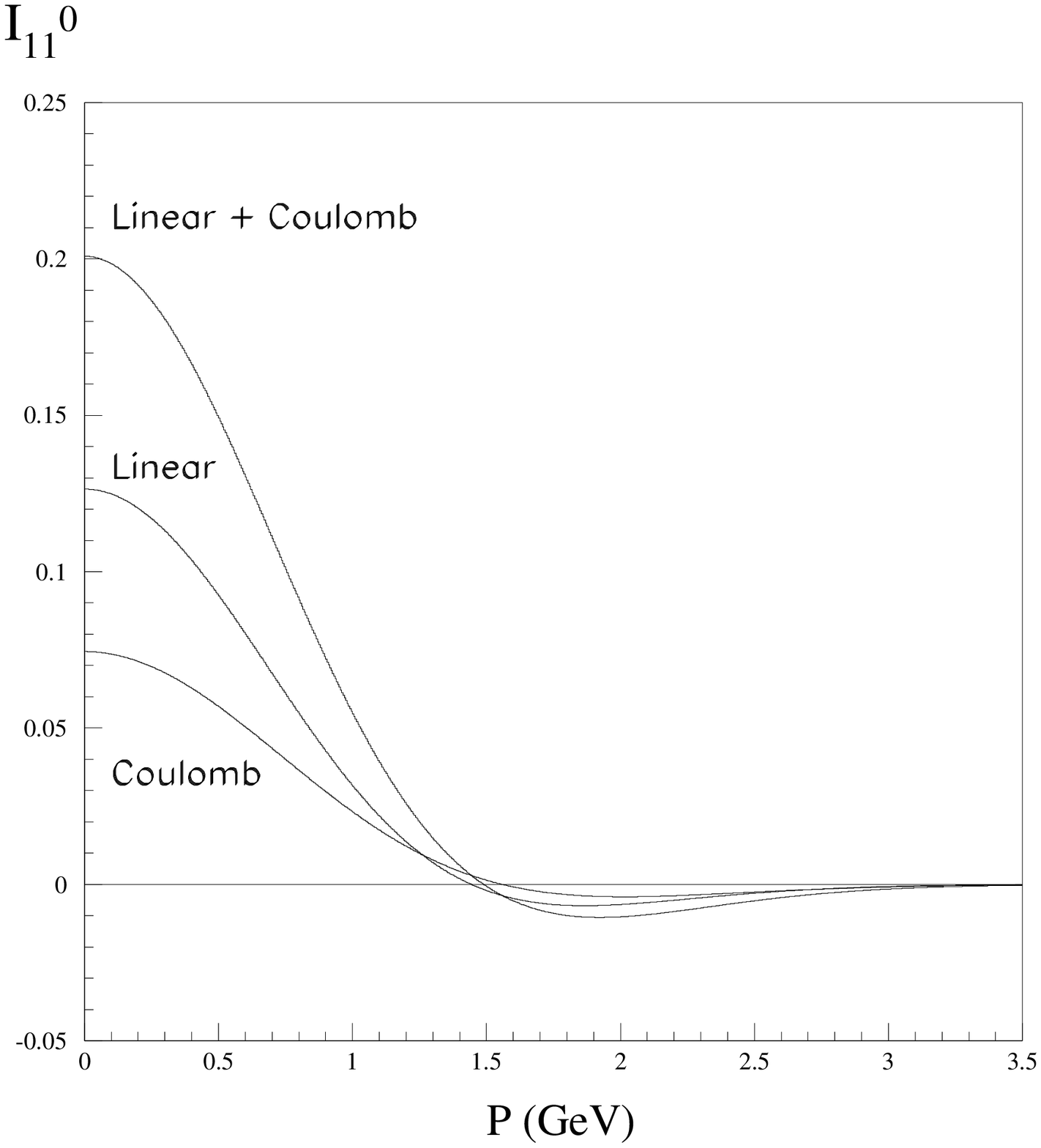,width=3.8in}
\caption[*]{$I_{11}^0(P)$ for the potential A with $\mu_b$=0.46 GeV.}
\label{fig:inlseta046l0}
\end{figure}

\begin{figure}
\centering
\epsfig{figure=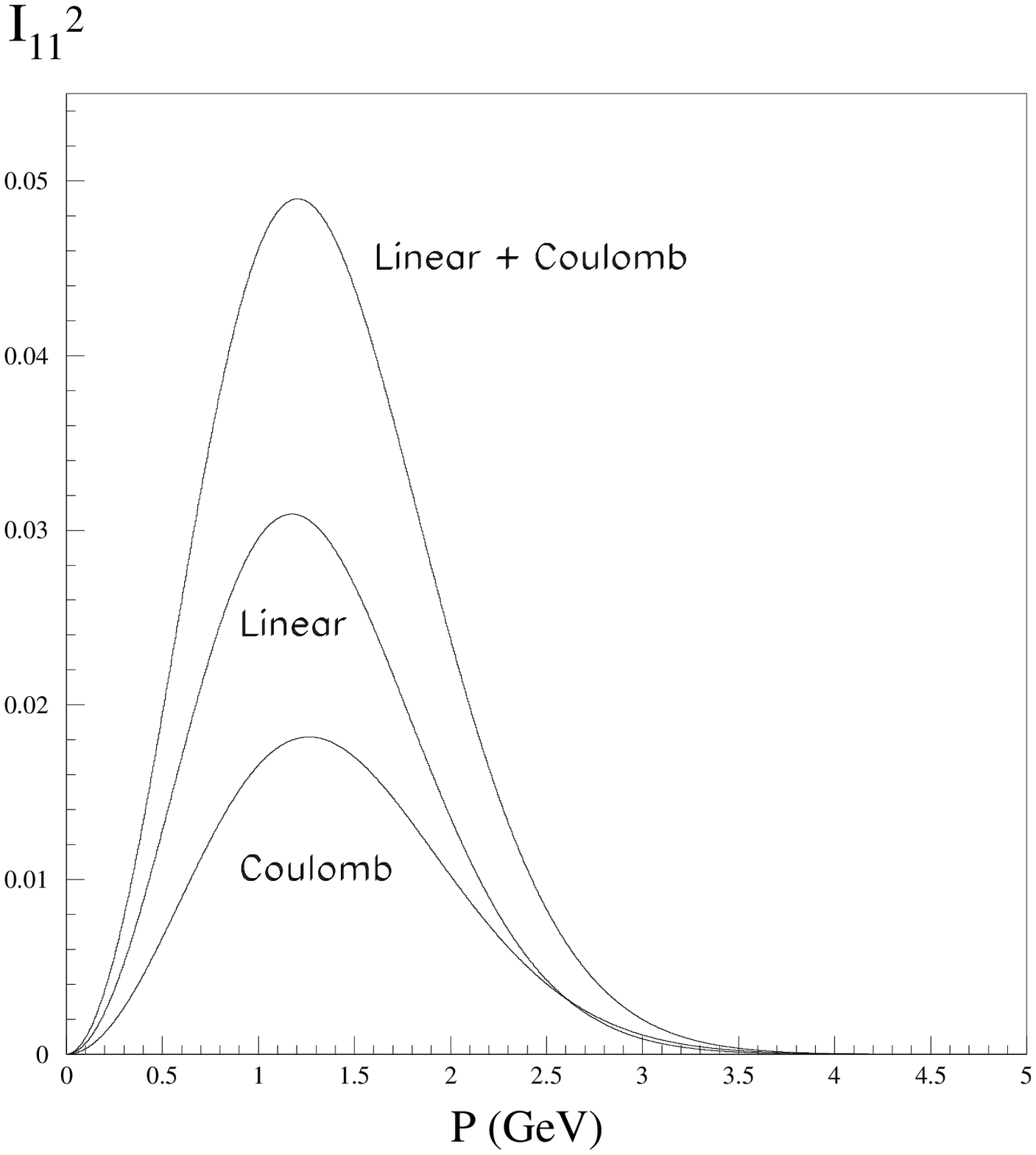,width=3.8in}
\caption[*]{$I_{11}^2(P)$ for the potential A with $\mu_b$=0.46 GeV.}
\label{fig:inlseta046l2}
\end{figure}

\begin{figure}
\centering
\epsfig{figure=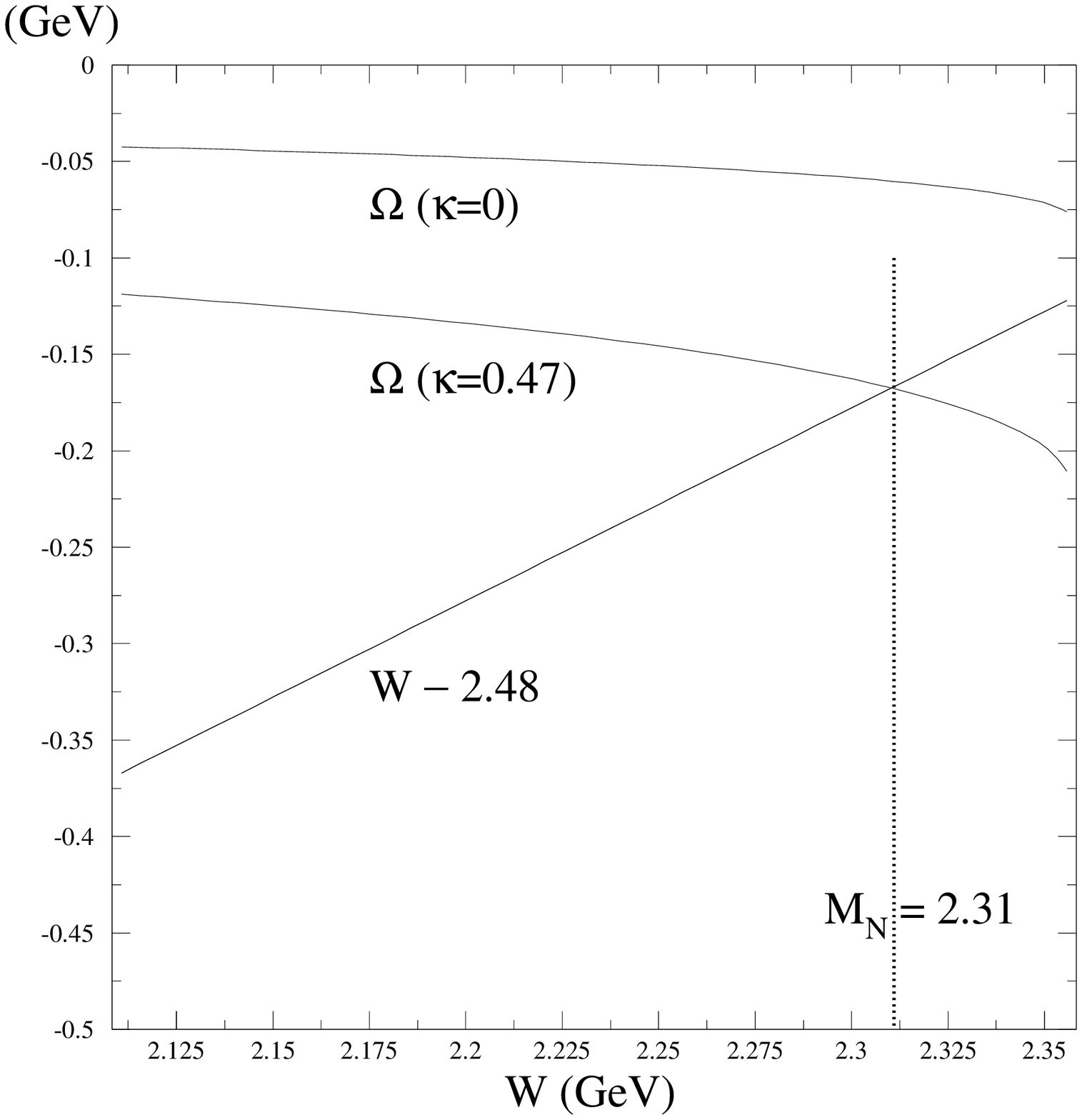,width=6.0in}
\caption{The result when the potential B and $\mu_b$= 0.46 GeV were used.
The upper first and second lines are $\Omega$ when the Coulombic
part is not included and when it is include, respectively, and the lower diagonal
line is $W-M^{\rm bare}$ where $M^{\rm bare}$=2.48 GeV given in \cite{IG}.
As we see here, $\Omega$ without the Coulombic part does not meet with
$W-M^{\rm bare}$, which means that there does not exist a solution of the
eigenvalue equation (\ref{a9}) below thresholds. On the other hand, $\Omega$ with
the Coulombic part meets with $W-M^{\rm bare}$ below thresholds and the value of
$W$=2.31 GeV at the meeting point is the eigenvalue $M_N$ of (\ref{a9}).}
\label{fig:setb046}
\end{figure}

For $V(r)$ in (\ref{a14}) we use the following two parameterizations in our
calculation.\\
(A) Eichten et al. \cite{eichten12}:
$\kappa$=0.517, $a$=2.12 GeV, $m_c$=1.84 GeV.\\
(B) Hagiwara et al. \cite{hagiwara}: 
$\kappa$=0.47, $a=1/{\sqrt{0.19}}$ GeV, $m_c$=1.32 GeV.\\
For the masses of $s$ and $u,d$ quarks, we use
$m_s$=0.55 GeV, $m_{u,d}$=0.33 GeV.

Following \cite{IG},
we use the radial wave functions of $c{\bar{u}}$, $c{\bar{d}}$,
${\bar{s}}u$, ${\bar{s}}d$ states with $L=0$ which are given
by the ground state harmonic-oscillator wave functions
$(2{\mu_a}^{3\over 2}/{\pi}^{1\over 4})$
${\rm exp}(-{\mu_a}^2r^2/2)$, and that of $c{\bar{s}}$
state with $L=1$ by the first excited harmonic-oscillator wave function
$(8^{1\over 2}{\mu_b}^{5\over 2}/3^{1\over 2}{\pi}^{1\over 4})\, r\,
{\rm exp}(-{\mu_b}^2r^2/2)$.
$\beta$ in (\ref{a17})
and $\mu_a$ here are related by $\beta = \mu_a^2/2$.
For the value of $\beta$ in the calculations of (\ref{a17}) and (\ref{a18}),
we take $\mu_a$=0.61 GeV which is the average value of
$\mu_a(K)$=0.61 GeV and
$\mu_a(D)$=0.60 GeV given in \cite{BCDGS}.
For the value of $\rho_Q$ in (\ref{a18}) defined as $m_Q/(m_q+m_Q)$, we take
$m_Q$ as $(m_c+m_s)/2$.
The formula for $I_{nL}^l(P)$ in (\ref{a18}) was derived for the case where two
mesons in the coupled channel have the same heavy quark such as $DD$ \cite{eichten12}.
However, in our present study two mesons are $D$ and $K$ mesons, and (\ref{a18})
is not exact in our system. So, we approximate our system
by using the formula for $I_{nL}^l(P)$ given in (\ref{a18}) with the above-mentioned
values of $\beta$ and $\rho_Q$.
For the value ${\mu_b}$ which comes into the $c{\bar{s}}$ state
radial wave function $R_{nL}(r)$
in (\ref{a18}), we take two prototype values:
${\mu_b}$=0.46 GeV given in \cite{BCDGS} for $D_{sJ}(2632)$ and
${\mu_b}$=0.57  GeV in \cite{GK} for ${}^3P_0(c{\bar{s}})$.
In the calculation in this paper we also adopt the approximation of
taking only one $c{\bar{s}}$ bound state of $1{}^3P_0$ for the
quark-antiquark bound state sector, whereas we take all the two-meson
continuum sector of $DK$, $D^*K$, $DK^*$ and $D^*K^*$.
That is, we do not include in our calculation the higher $c{\bar{s}}$ bound states
of $n{}^3P_0$ whose masses are located above the threshold.
If we include those higher $c{\bar{s}}$ bound states
in the calculation, the physical mass eigenvalue would be modified.
However, we expect the modification is not large since the $1{}^3P_0$ is
very close to the threshold and then the coupling channel effect to this state
is dominant.
There is also a merit of this approximation that one can see clearly how
the coupled channel effect gives rise to the mass shift.

The statistical coupling coefficient $C(Js,LL',J_1J_2,l)$ in (\ref{a16})
is 1 for $l=0$ in the $DK$ channel,
${1/3}$ for $l=0$ and ${8/3}$ for $l=2$ in the
$D^*K^*$ channel, and other coefficients are zero.
In the calculation we use the meson masses given in Table 1 \cite{pdg04}.
We calculate $I_{nL}^l(P)$ using (\ref{a18}).
The results of $I_{nL}^l(P)$ for $n=L=1$ and $l=0,2$ are presented in
Figs. \ref{fig:inlseta046l0} and \ref{fig:inlseta046l2}.
As we see in the figures, the contribution to $I_{nL}^l(P)$ from the Coulombic
part is about 60 \% of that from the linear part.

Ref. \cite{eichten12} ignored the Coulombic part of $V(r)$ in (\ref{a13}) in order
to simplify their calculations. They argued that this is justified because
small quark separations are not important in hadronic decays.
When Eichten et al. \cite{ELQ} analyzed the recently found $X(3872)$ state using
the coupled channel effect, they also did not include the Coulombic part in
(\ref{a13}) in their calculation by considering that it is a good approximation
to neglect all effects of the Coulomb piece in (\ref{a13}).
It has been a common practice to ignore the Coulombic part in (\ref{a13}).
However, it was first found by Zambetakis \cite{zam1,byers94} that if one includes the
Coulombic part in (\ref{a13}), he finds a significant modification of $I_{nL}^l$.
In fact, we find in this paper that the contribution to $I_{nL}^l$ from the Coulombic
part is about 60 \% of that from the linear part, and then the magnitude of
$\Omega$ resulting when both linear and Coulombic parts are included is about 2.6 
times that resulting when only the linear part is considered.

The fact that the inclusion of the Coulombic part in (\ref{a13}) increases
the magnitude of $\Omega$ about 2.6 times is very important in the analysis
of this paper.
Fig. \ref{fig:setb046}
is the result which we get when the potential B and $\mu_b$= 0.46 GeV are used.
The upper first and second lines are $\Omega$ when the Coulombic
part is not included (that is, when only the linear part is considered)
and when it is included, respectively, and the lower diagonal
line is $W-M^{\rm bare}$ where $M^{\rm bare}$=2.48 GeV given in \cite{IG}.
As we see the figure, $\Omega$ without the Coulombic part does not meet with
$W-M^{\rm bare}$, which means that there does not exist a solution of the
eigenvalue equation (\ref{a9}) below thresholds. On the other hand, $\Omega$ with
the Coulombic part meets with $W-M^{\rm bare}$ below thresholds and the value of
$W$=2.31 GeV at the meeting point is the eigenvalue $M_N$ of (\ref{a9}).
The experimentally measured value of $D_{sJ}^*(2317)$ mass is 2317.4$\pm$0.9 MeV,
which is 40.9$\pm$1.0 MeV below the lowest threshold energy 2358$\pm$0.5 MeV
of $D^0K^+$.
For all the cases of the parameterizations given in Table 2,
we have similar situations as that shown in Fig. \ref{fig:setb046}.
In Table 2 we present the physical mass eigenvalue $M_N$ and the mass shift
$\Delta M_N \equiv M_N - M^{\rm bare}$(2.48 GeV).
The values of $Z$ which we calculated using (\ref{a11}) are also presented
in the table.

\begingroup
%\squeezetable
\begin{table}[t]
%\vspace*{0.3cm}
\label{tab:eigenvalue}
\begin{center}
\begin{tabular}{|l|l|c|c|c|}
\hline\hline
\multicolumn{2}{|c|}{Parameterizations} & $M_N$ & $\Delta M_N$ & $Z$ \\
\cline{1-2}
\ \ Potential \ \ & \ \ \ $\mu_b$ (GeV) \ \ \ & \ \ \ \ (GeV) \ \ \ \
& \ \ \ \ (GeV) \ \ \ \ &\ \ \ \ \ \ \ \ \ \ \ \ \ \ \\
\hline\hline
\ \ \ \ \ \ A \ \ & \ \ \ \ \ 0.46             & 2.29 & --0.19 & 0.67 \\
\ \ \ \ \ \ \ \ \ & \ \ \ \ \ 0.57             & 2.27 & --0.21 & 0.70 \\
\hline
\ \ \ \ \ \ B \ \ & \ \ \ \ \ 0.46             & 2.31 & --0.17 & 0.67 \\
\ \ \ \ \ \ \ \ \ & \ \ \ \ \ 0.57             & 2.29 & --0.19 & 0.71 \\
\hline\hline
\end{tabular}
\end{center}
\vspace*{-0.5cm}
\caption{The results of $M_N$, $\Delta M_N$ and $Z$ for each parameterization of
the potential and the $\mu_b$ value, when $\rho_Q(\equiv m_Q/(m_Q+m_{u,d}))$ with
$m_Q=(m_c+m_s)/2$ was used.}
\end{table}
\endgroup

\begingroup
%\squeezetable
\begin{table}[t]
%\vspace*{0.3cm}
\label{tab:eigenvalue}
\begin{center}
\begin{tabular}{|l|l|l|c|c|c|}
\hline\hline
\multicolumn{3}{|c|}{Parameterizations} & $M_N$ & $\Delta M_N$ & $Z$ \\
\cline{1-3}
\ \ \ \ \ $\rho_Q$\ \ \ \ \ & \ \ Potential \ \ & \ \ \ $\mu_b$ (GeV) \ \ \ &
\ \ \ \ (GeV) \ \ \ \
& \ \ \ \ (GeV) \ \ \ \ &\ \ \ \ \ \ \ \ \ \ \ \ \ \ \\
\hline\hline
   &
\ \ \ \ \ \ A \ \ & \ \ \ \ \ 0.46             & 2.31 & --0.17 & 0.63 \\
\ \ \ \ \ $\rho_c$\ \ \ \ \ &
\ \ \ \ \ \ \ \ \ & \ \ \ \ \ 0.57             & 2.29 & --0.19 & 0.68 \\
\cline{2-6}
   &
\ \ \ \ \ \ B \ \ & \ \ \ \ \ 0.46             & 2.32 & --0.16 & 0.62 \\
   &
\ \ \ \ \ \ \ \ \ & \ \ \ \ \ 0.57             & 2.31 & --0.17 & 0.67 \\
\hline\hline
   &
\ \ \ \ \ \ A \ \ & \ \ \ \ \ 0.46             & 2.23 & --0.25 & 0.73 \\
\ \ \ \ \ $\rho_s$\ \ \ \ \ &
\ \ \ \ \ \ \ \ \ & \ \ \ \ \ 0.57             & 2.19 & --0.29 & 0.75 \\
\cline{2-6}
   &
\ \ \ \ \ \ B \ \ & \ \ \ \ \ 0.46             & 2.27 & --0.21 & 0.73 \\
   &
\ \ \ \ \ \ \ \ \ & \ \ \ \ \ 0.57             & 2.25 & --0.23 & 0.75 \\
\hline\hline
\end{tabular}
\end{center}
\vspace*{-0.5cm}
\caption{The results of $M_N$, $\Delta M_N$ and $Z$ for each parameterization of 
the potential and the $\mu_b$ value, when $\rho_Q(\equiv m_Q/(m_Q+m_{u,d}))$ with 
$m_Q=m_c$ or $m_s$ was used in order to see the sensitivity of the results to
the parameter values.}
\end{table}
\endgroup

In order to see the sensitivity of the results to the value of the parameter
$\rho_Q$, we also performed the same calculation for $\rho_Q=\rho_c$ and
$\rho_s$, where $m_Q=m_c$ and $m_s$ in $\rho_Q\equiv m_Q/(m_Q+m_{u,d})$,
and the results are presented in Table 3.
For all these different parameterizations, it is still true that
the contribution to $I_{nL}^l$ from the Coulombic
part is about 60 \% of that from the linear part, and then the magnitude of
$\Omega$ resulting when both linear and Coulombic parts are included is about 2.6
times that resulting when only the linear part is considered.
Among all the cases in Table 2 and 3, only for the cases in the 5th and 6th rows
of Table 3 there exists an eigenstate also for the $\Omega$ with $\kappa = 0$,
with its eigenvalue $M_N$ which is just belw the threshold of $D^0K^+$.
However, the parameterizations of those cases are not physically acceptable
since their mass eigenvalues presented in the 5th and 6th rows of
Table 3 are too small compared to the
measured value.

\section{Conclusion}

It has been a common practice to ignore the Coulombic part in the calculation
of the transition amplitude between two-quark bound and two-meson continuum states.
However, the calculation in this paper shows explicitly that it is about 60 \%
larger when both the linear and Coulombic parts are included in the calculation
compared to the case where the latter part is ignored.
Consequently, the magnitude of $\Omega$ becomes about 2.6 times larger when
the Coulombic part is also included.
This fact is crucial in the analysis of this paper:
When the Coulombic part is not included, there is no physical mass eigenstate
below the threshold of $D^0K^+$ for physically acceptable parameterizations.
On the other hand, when the Coulombic part is included, there exists a
physical mass eigenstate with its mass close to the measured $D_{sJ}^*(2317)$
mass in a natural manner.
Therefore, it was shown that the coupled channel effect explains why the mass of
$D_{sJ}^*$(2317) is about 160 to 170 MeV lower than the bare mass of the
$1{}^3P_0$ bound state obtained from the most potential model calculations.

%%%%%%%%%%%%%%%%%%%%%%%%%%%%%%%%%%%%%%
\section*{Acknowledgments}
%%%%%%%%%%%%%%%%%%%%%%%%%%%%%%%%%%%%%%
One of the authors (D.S.H.) wishes to thank Nina Byers and Ted Barnes
for enlightening discussions.
This work was supported in part by the International Cooperation Program of
the KISTEP (Korea Institute of Science \& Technology Evaluation and Planning).

\end{document}